\documentclass{article}
\usepackage{spconf,amsmath,graphicx}
\usepackage{xcolor}
\usepackage{diagbox}
\usepackage{multirow}
\usepackage{booktabs}
\usepackage{subfigure}
\usepackage{mathtools}
\usepackage{amsmath}
\usepackage{cite}

\title{Expressive Voice Conversion: \\ A Joint Framework for Speaker Identity and Emotional Style Transfer}
\name{Zongyang Du$^{1,2}$\thanks{\textbf{Speech Samples:} https://zy-du.github.io/ASRU21/}, Berrak Sisman$^1$, Kun Zhou$^2$, Haizhou Li$^{2,3}$}
\address{
  $^1$Singapore University of Technology and Design, Singapore\\
  $^2$National University of Singapore, Singapore \\
  $^3$The Chinese University of Hong Kong (Shenzhen), China
  }

\begin{document}

%
\maketitle
\begin{abstract}
Traditional voice conversion (VC) has been focused on speaker identity conversion for speech with a neutral expression. We note that emotional expression plays an essential role in daily communication, and the emotional style of speech can be speaker-dependent. 
In this paper, we study a technique to jointly convert the speaker identity and speaker-dependent emotional style, that is called expressive voice conversion. 
We propose a StarGAN-based framework to learn a many-to-many mapping across different speakers, that takes into account speaker-dependent emotional style without the need for parallel data. To this end, we condition the generator on emotional style encoding derived from a  pre-trained speech emotion recognition (SER) model.
The experiments validate the effectiveness of our proposed framework in both objective and subjective evaluations. To our best knowledge, this is the first study on expressive voice conversion.
\end{abstract}
\section{Introduction}
\label{sec:intro}
\vspace{-2mm}
    
    


 Traditional voice conversion (VC) aims to modify one's voice to sound like that of another while keeping linguistic content and emotional style unchanged \cite{sisman2020overview}. VC is an enabling technology for various tasks, such as conversational assistants, cross-lingual speech synthesis \cite{zhou2019cross,sisman2019study,du2020spectrum}, and speaker verification \cite{wu2015spoofing}. In this   paper, we formulate a new research topic denoted as \textit{Expressive Voice Conversion}, and propose a solution that jointly performs speaker identity and emotional style transfer for emotional speakers.

It is known that human speech is expressive and emotive in nature, and people usually express themselves with various emotional states such as happy and sad \cite{scherer1991vocal}. Studies also reveal the fact that emotional speech style contains speaker-dependent elements which has never been studied in traditional voice conversion \cite{middleton1989emotional,arnold1960emotion,cohen2003emotional}. 
The study of expressive voice conversion is motivated to fill the gap. 

\begin{figure}[t]
    \centering
    \includegraphics[scale = 0.4]{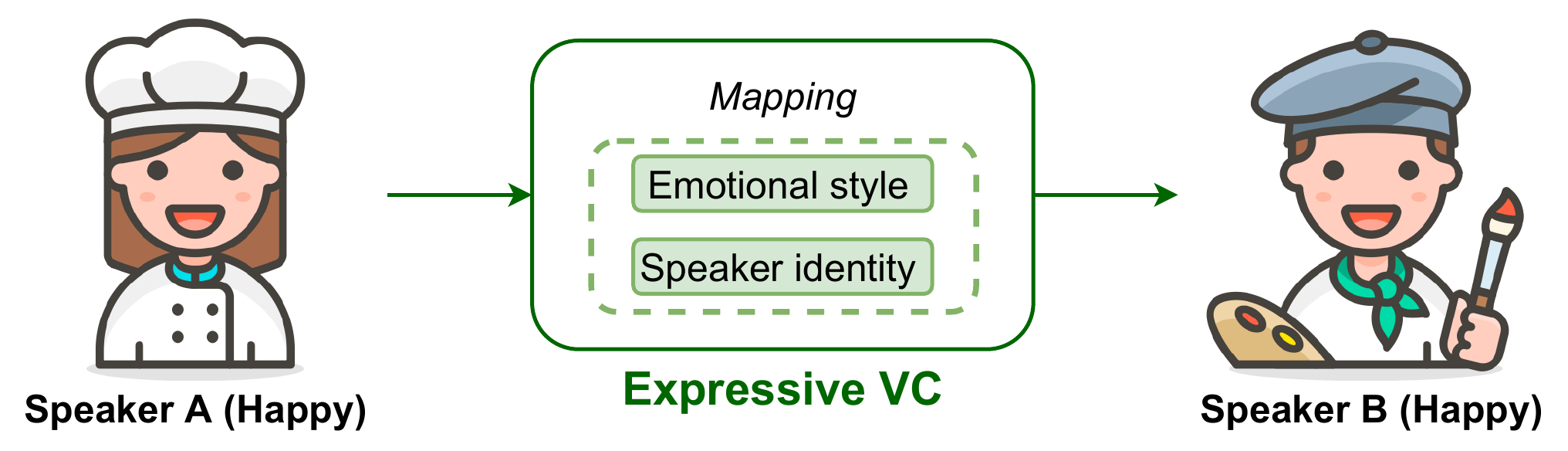}
    \vspace{-1mm}
    \caption{An example of expressive voice conversion which aims to convert the happy utterances expressed by speaker A to sound like those of speaker B without changing the speaking content. We note that an individual expresses the happiness in one's own style.}
    \vspace{-2mm}
    \label{fig:vc_emotions}
\end{figure}

In expressive voice conversion, both speaker identity and emotional style need to be carefully dealt with, as illustrated in Fig.\ref{fig:vc_emotions}. In general, emotional speech style is more complex and difficult to model \cite{ekman1992argument,hirschberg2004pragmatics,dai2009comparing} than speaker identity which can be characterized by frame-based spectral features.  It is known that emotional speech style can be expressed with universal patterns linked to lexical stress and speech acts (speaker-independent emotional style) \cite{kotti2012speaker,zhou2020converting}, and with personal and other background characteristics associated with an individual (speaker-dependent emotional style) \cite{arnold1960emotion,dai2009comparing}. 
It makes sense to only transfer the speaker-dependent emotional style while preserving the speaker-independent elements in the process, which makes expressive voice conversion a challenging task.

While expressive voice conversion shares similarities with emotional voice conversion, it differs from emotional voice conversion in many ways. For example, emotional voice conversion aims to convert the emotional state of an utterance while preserving the speaker identity \cite{robinson2019sequence,zhou2021emotional}.
In emotional voice conversion, the emotional state similarity (e.g. happy, sad, angry) between generated and target speech is the major concern \cite{Zhou2020,gao2019nonparallel,zhou2021vaw}. However, in expressive voice conversion, we do not seek to change the emotional state, instead we jointly transfer speaker-dependent characteristics such as speaker identity and emotional style to the converted speech.

We note that some previous studies \cite{inproceedings,hsia2007conversion} refer expressive voice conversion to  emotional voice conversion. To avoid any confusion, in this paper we refer  expressive voice conversion to converting an input speech towards a target speaker both in terms of speaker identity and emotional style. We propose an expressive voice conversion framework that performs the conversion without the need for parallel data. 


In traditional voice conversion, we seek to convert the speaker identity while preserving the linguistic information. As speaker identity is determined by the vocal timbre that is manifested in spectrum \cite{ramakrishnan2012speech}, early voice conversion studies are mainly focused on modeling the spectral mapping between the source and target with statistical parametric methods such as Gaussian mixture model (GMM)~\cite{mashimo2002cross,toda2007voice,6694179}.  
 Recent deep learning methods such as deep neural network (DNN) \cite{hinton2006fast}, and recurrent neural network (RNN) \cite{nakashika2014high} 
have significantly improved the performance. We note that these methods require parallel data, that limits the scope of  real-life applications.  To enable non-parallel voice conversion, many frameworks based on generative models such as variational auto-encoders (VAE) \cite{hsu2016voice,van2020vector,qian2020f0, huang2018voice} and generative adversarial network (GAN)\cite {hsu2017voice,ferro2021cyclegan,sisman2018adaptive,kaneko2018cyclegan,kaneko2019cyclegan,kaneko2020cyclegan,Kaneko2017ParallelDataFreeVC} are proposed in recent years. Among the GAN-based methods, StarGAN-VC \cite{kameoka2018stargan,kaneko2019stargan} extends non-parallel VC from two domains to multiple domains. It allows for sharing of knowledge across multiple pairwise mapping, and represents one of the successful attempts in non-parallel VC, which motivates our study.   We would like to study the use of a StarGAN architecture to learn a translation model across multiple spectral domains from different speakers with different emotional styles. 

Our proposed framework consists of two stages for training: 1) emotional style descriptor training, where we train a SER network to learn emotion-related information such as emotional style features through acoustic features; 2) StarGAN training, where we condition the generator with both speaker label and emotional style features.
We believe that emotional style features serve as an excellent tool to describe the emotional style in a continuous space, thus more suitable for speaker-dependent emotional style transfer. 



The main contributions of this paper include: 
1) we formulate a novel research topic, expressive voice conversion, as an extension of voice conversion research;
2) we propose an expressive voice conversion framework based on StarGAN without the need for parallel data, and that is flexible for many-to-many conversion; 
3) we conduct experiments on a publicly available multi-speaker database under different emotional states, and show the effectiveness of our proposed expressive voice conversion framework for both emotional style and speaker identity transfer. To our best knowledge, this is the first paper to study expressive voice conversion.

The rest of this paper is organized as follows: In Section 2, we introduce the related work. In Section 3, we investigate the speaker-dependent emotional style with deep emotional style features, which motivates our study. In Section 4, we introduce our proposed voice conversion framework with StarGAN. In Section 4, we report the experiments. Section 5 concludes the study.

\section{Related Work}
\vspace{-2mm}
\label{RW}

\subsection{Traditional voice conversion and datasets}
Traditional voice conversion is studied for speech with a neutral expression, where  voice quality has been the main focus. The most widely used speech databases for voice conversion include VCTK database \cite{yamagishi2019cstr}, CMU-Arctic database \cite{kominek2004cmu}, and Voice Conversion Challenge (VCC) corpus \cite{toda2016voice,lorenzo2018voice,zhao2020voice}. Since these speech databases are emotion-free, it is straightforward to pay attention to the vocal timbre when conducting voice conversion research. 

We note that voice conversion with emotional speech data has never been studied before, mostly due to the lack of suitable databases. Recently, ESD database \cite{zhou2021emotional} is released for emotional voice conversion studies, which includes multi-speaker and multi-lingual parallel speech data with different emotions, and makes possible emotion-related studies, such as expressive voice conversion.

\begin{figure*}
\centering
\subfigure[The Euclidean distance and RMSE from Speaker 0013 to all other speakers]{
\begin{minipage}[t]{0.3\linewidth}
\centering
\includegraphics[width=5cm]{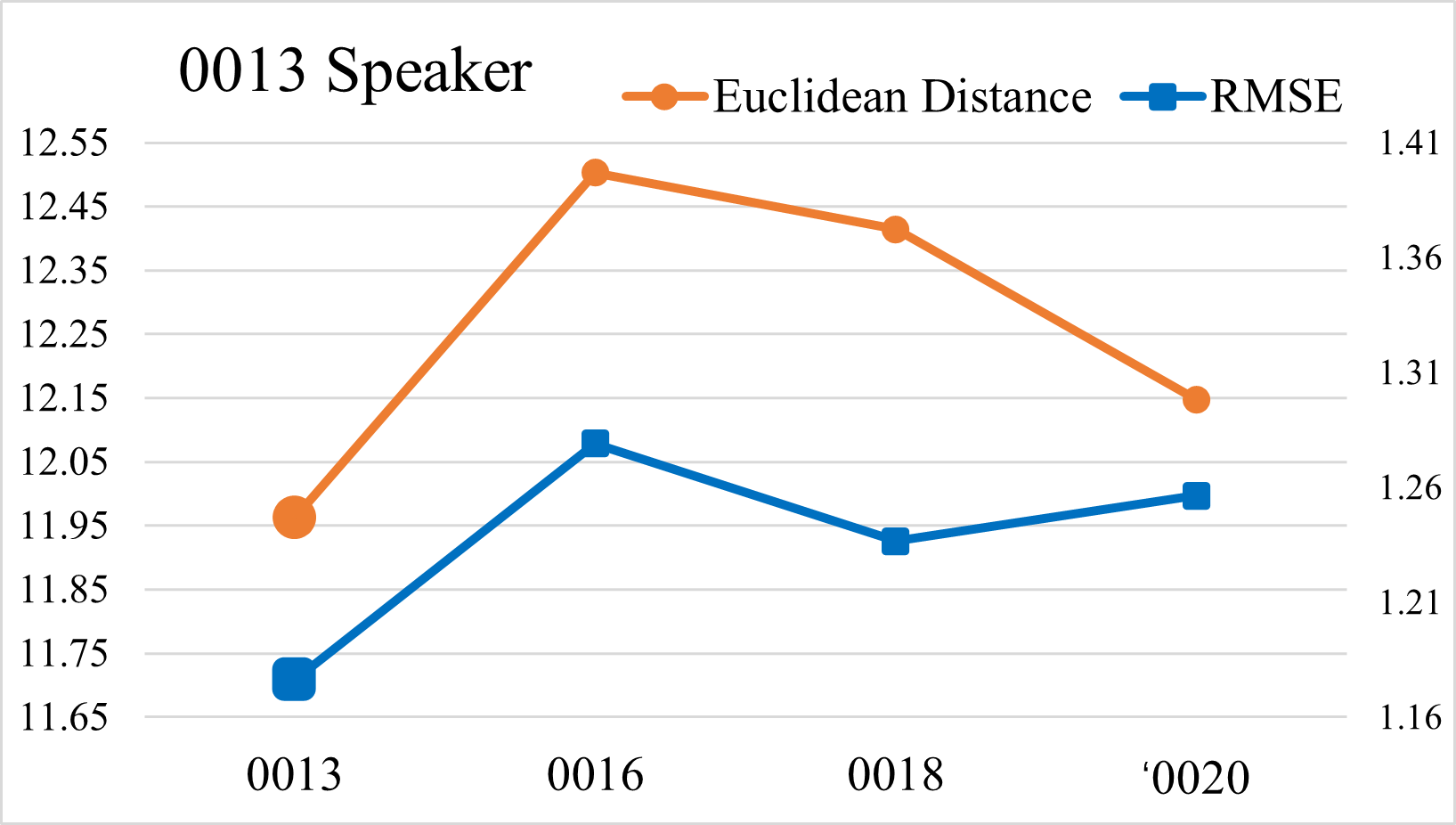}
\end{minipage}%
}\hspace{5mm}
\subfigure [The Euclidean distance and RMSE from Speaker 0016 to all other speakers]{
\begin{minipage}[t]{0.3\linewidth}
\centering
\includegraphics[width=5cm]{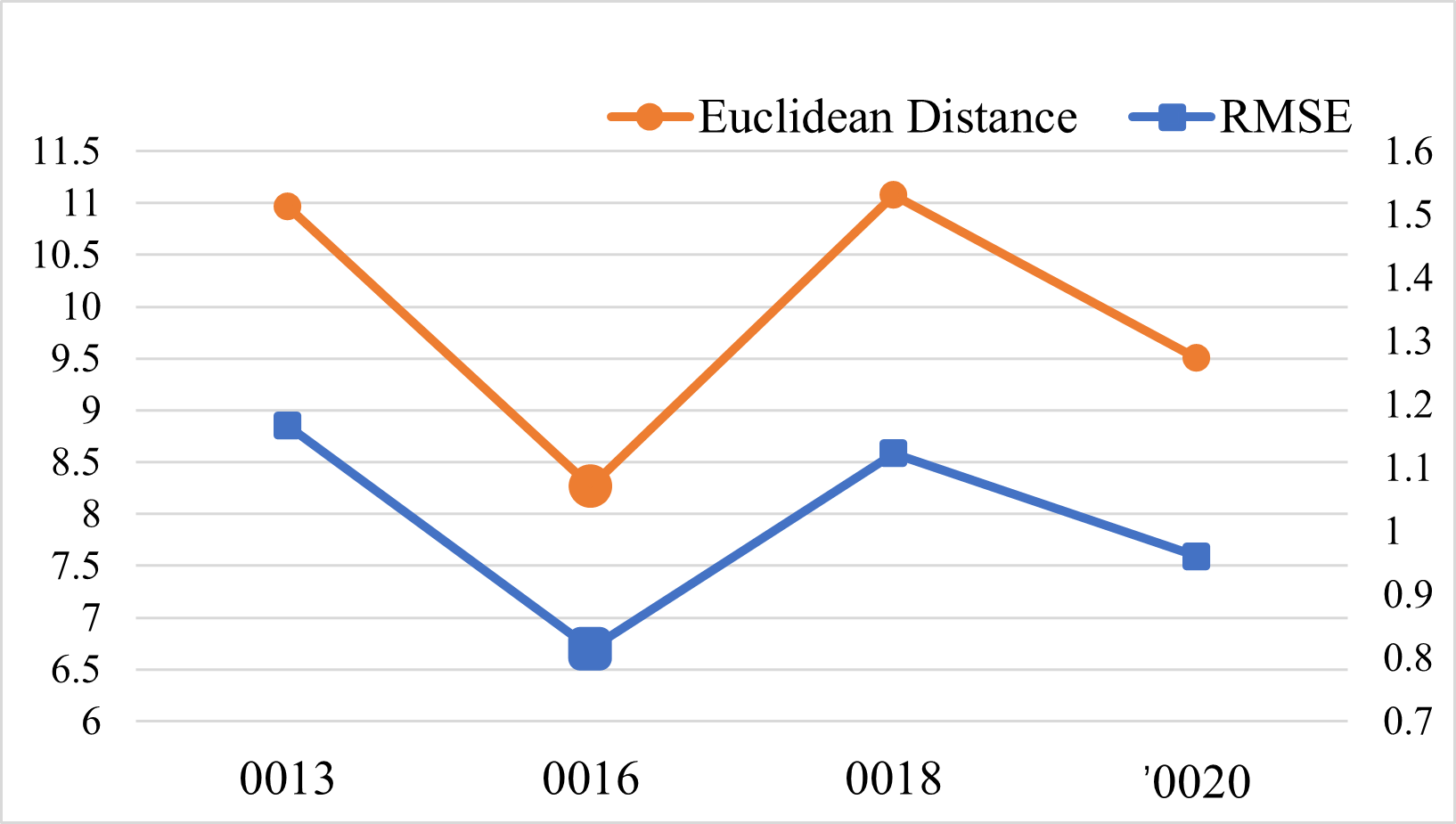}
\end{minipage}%
}\hspace{5mm}
\subfigure[The Euclidean distance and RMSE from Speaker 0018 to all other speakers]{
\begin{minipage}[t]{0.3\linewidth}
\centering
\includegraphics[width=5cm]{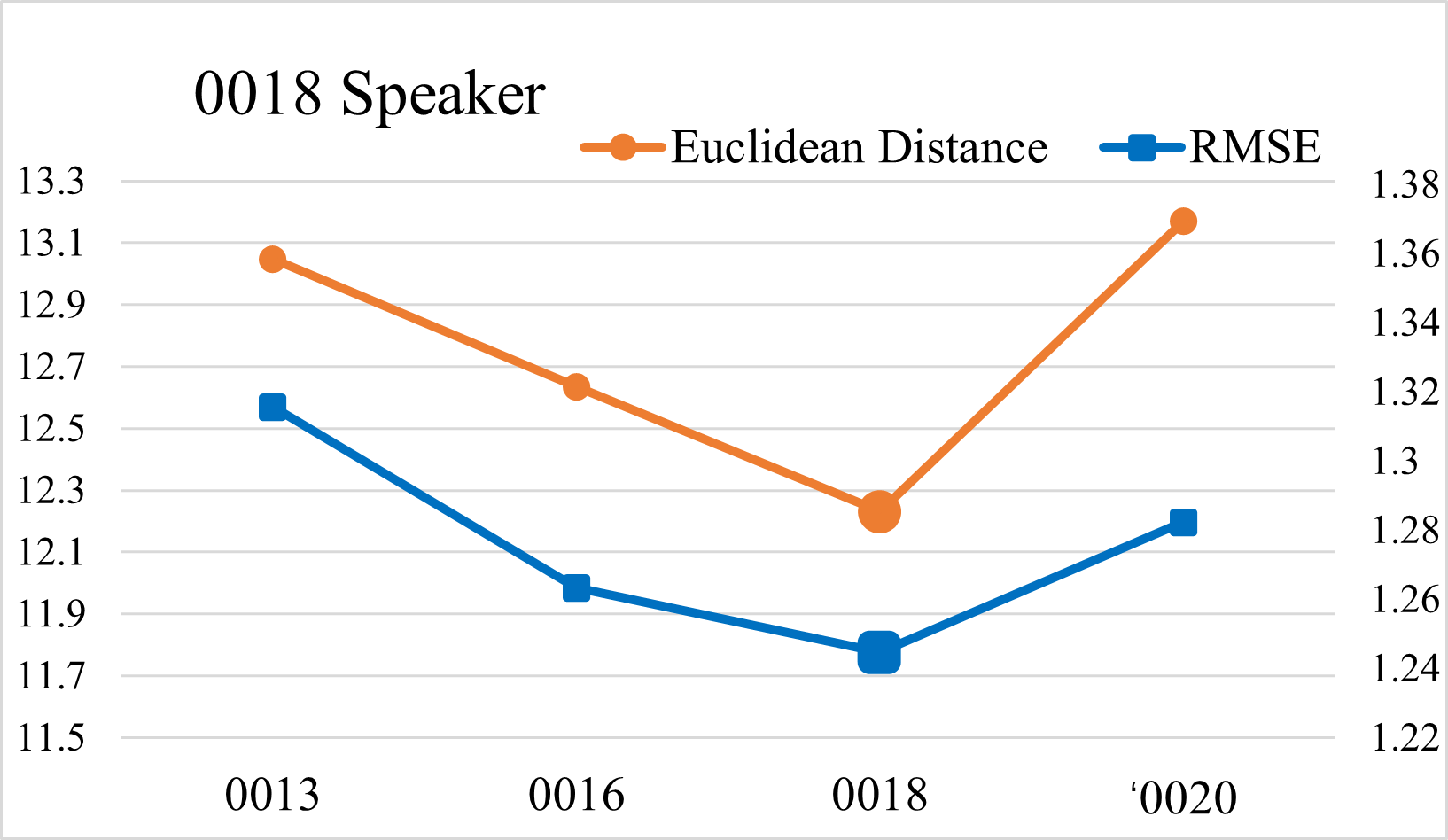}
\end{minipage}
}\hspace{5mm}
\vspace{-5mm}
\centering
\caption{RMSE and Euclidean distance calculation results between emotional style features for 50 utterances from different speaker-pairs in sad emotion state.}
\label{fig:unique}
\vspace{-4mm}
\end{figure*}

\subsection{StarGAN for voice conversion}
\vspace{-2mm}
 
 StarGAN \cite{choi2018stargan} is first proposed to multi-domain image-to-image translation in computer vision, and then adopted to voice conversion \cite{kameoka2018stargan}. StarGAN \cite{kameoka2018stargan} and its variants \cite{kaneko2019stargan,kaneko2020cyclegan,ferro2021cyclegan} have shown promising results in many-to-many voice conversion without the need for parallel data. A StarGAN consists of a generator, a discriminator, and a domain classifier. The generator takes both spectral features and domain information such as one-hot label, and translates the spectral features into the corresponding domain while the discriminator distinguishes whether the input features are real or fake. A domain classifier is used to further verify the label correctness of both real and generated spectral features. StarGAN also has been applied to other tasks such as affect generation \cite{kollias2020va} and emotional voice conversion \cite{9054579}. Motivated by the success of StarGAN in many-to-many voice conversion, we extend the idea and propose a many-to-many VC framework that can jointly transfer both speaker identity and emotional style, which will be further introduced in Section 4.

\section{Speaker-dependent Emotional Style}
\vspace{-2mm}


Emotional feature extraction has been a research hotspot in speech emotion recognition (SER) \cite{akccay2020speech}. With the advent of deep learning, there has been a shift from traditional human-crafted emotional features such as those extracted by low level descriptors (LLDs) \cite{wimmer2008low} or openSMILE \cite{eyben2010opensmile}, to the features automatically learned by deep neural networks (DNN) \cite{schuller2020review}. Many studies \cite{schuller2020review,latif2020deep} have shown that DNNs are capable of extracting hierarchical feature representations from expressive speech, which are more suitable for SER.
Meanwhile, recent speech synthesis studies \cite{um2020emotional,liu2021expressive, zhou2021seen} also propose to leverage those deep emotional features to characterize different emotional styles over a continuum \cite{kim2019dnn}.
These successful attempts have served as the source of motivation for this paper.


\begin{figure}[t]
\vspace{-3mm}
\centering
\label{tsne}
\subfigure[Speaker 0013 (M)]{
\begin{minipage}[t]{0.5\linewidth}
\centering
\includegraphics[width=4.2cm]{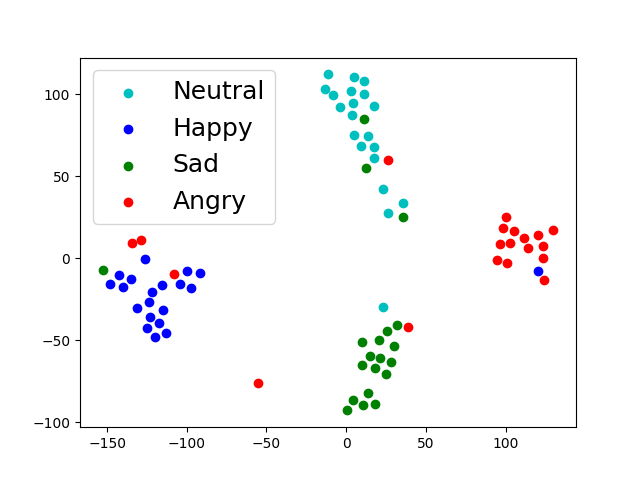}
\end{minipage}%
}%
\subfigure[Speaker 0016 (F)]{
\begin{minipage}[t]{0.5\linewidth}
\centering
\includegraphics[width=4.2cm]{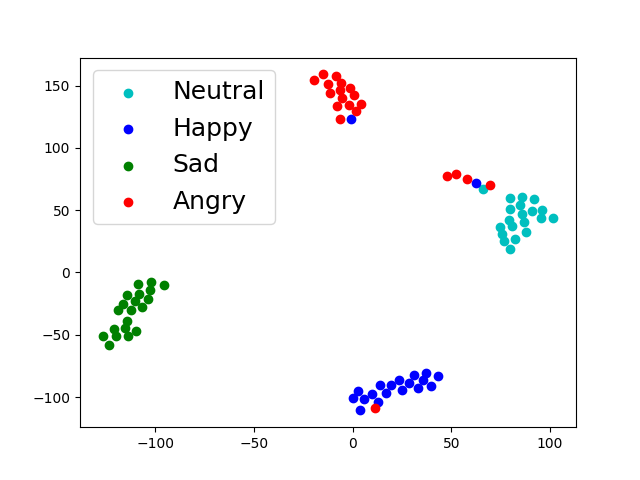}
\end{minipage}%
}%

\vspace{-3mm}
\subfigure[Speaker 0018 (F)]{
\begin{minipage}[t]{0.5\linewidth}
\flushleft
\includegraphics[width=4.2cm]{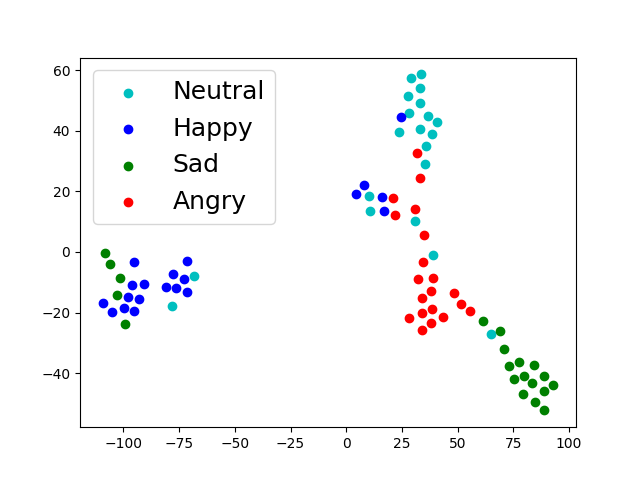}
\end{minipage}
}%
\subfigure[Speaker 0020 (M)]{
\begin{minipage}[t]{0.5\linewidth}
\centering
\includegraphics[width=4.2cm]{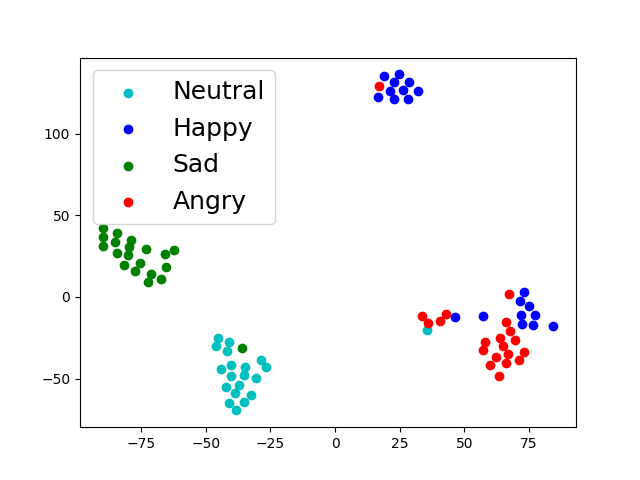}
\end{minipage}
}%

\centering
\vspace{-2mm}
\caption{T-SNE plot of emotional style features for 20 utterances with the same content but different emotions from 4 speakers.}
\vspace{-4mm}
\end{figure}

We propose to study a scenario for expressive voice conversion, where source and target speakers are expressing the same emotion in their own styles, that we refer to as speaker-dependent emotional styles. As emotional styles are hierarchical in nature, they are difficult to describe. We believe that deep emotional features bear a huge potential to describe both speaker-dependent and speaker-independent attributes for emotional styles. Therefore, we would like to investigate the use of deep emotional features for expressive voice conversion. 

\begin{figure*}[t]
    \centering
    \subfigure[Illustration of emotional style descriptor]{
        \begin{minipage}[t]{0.35\linewidth}

        \centering
        \includegraphics[scale = 0.35]{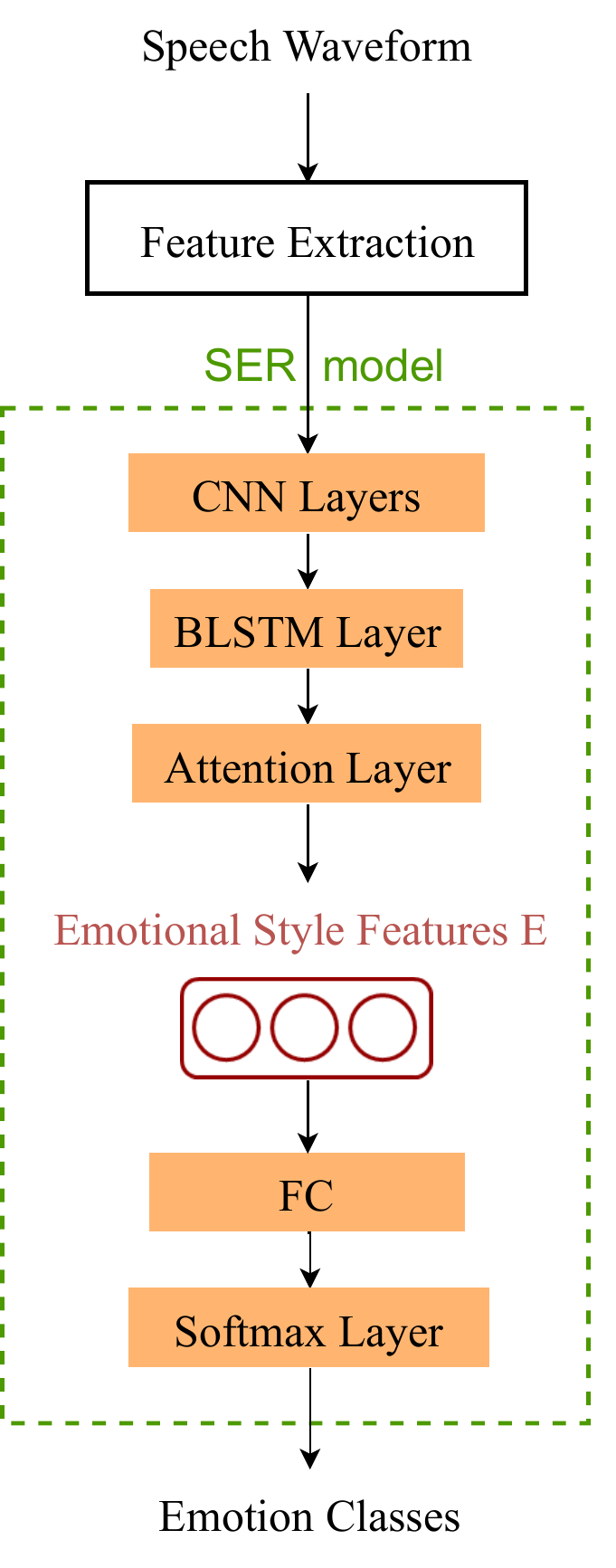}
        \label{SER}
        \end{minipage}
    }%
    \subfigure[JES-StarGAN training]{
        \begin{minipage}[t]{0.65\linewidth}
        \centering
        \includegraphics[scale = 0.35]{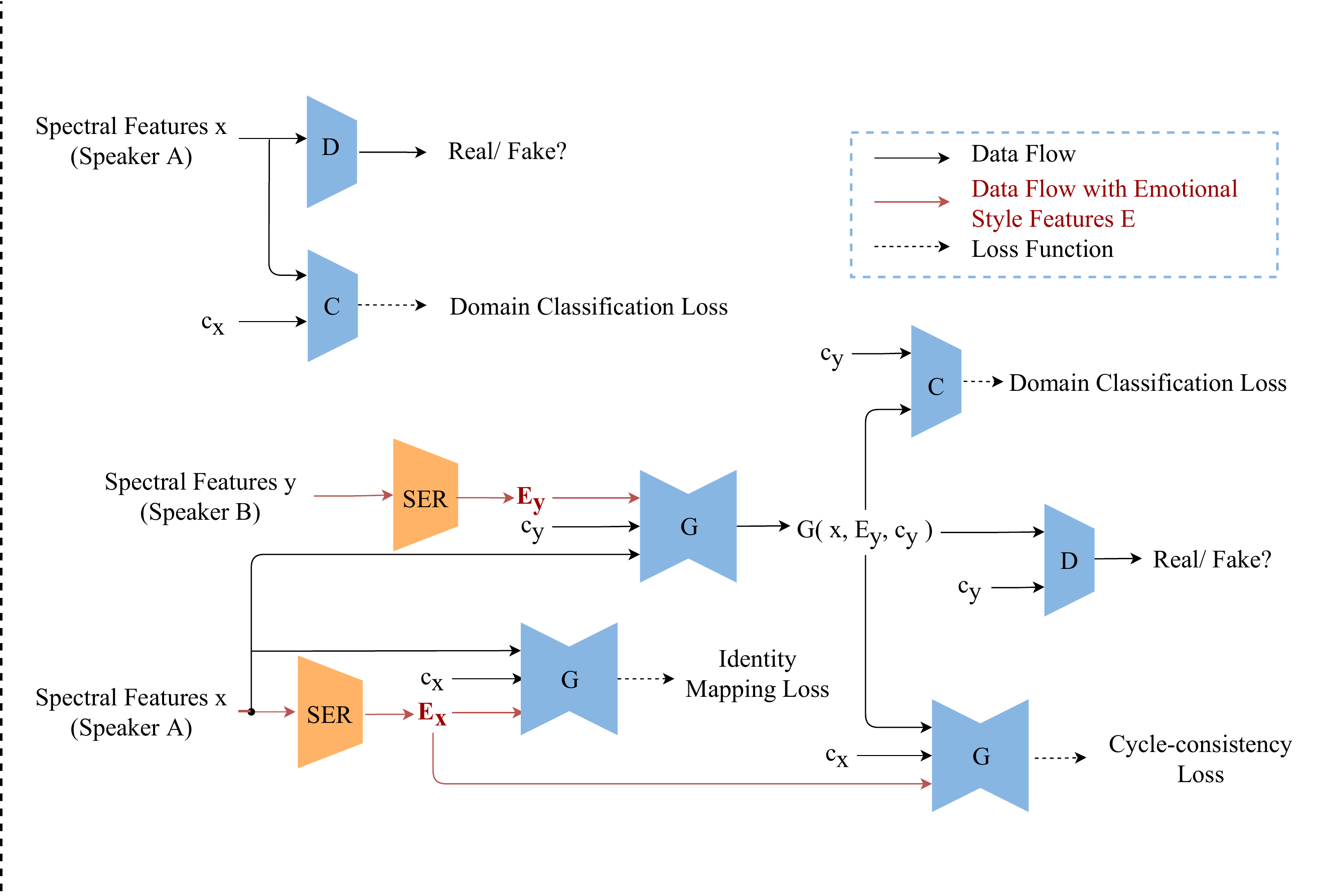}
        \label{StarGAN}
        \end{minipage}
    }%
    \vspace{-3mm}
    \caption{Schematic diagram of the proposed framework JES-StarGAN. Blue boxes represent the modules involved in the training and the yellow boxes represent the pre-trained modules. }
    \label{fig:training}
    \vspace{-3mm}
\end{figure*}

We first train a SER network and use the emotional style features before the last projection layer as illustrated in Fig.\ref{SER}.
We then use t-SNE algorithm \cite{van2008visualizing} to visualize the emotional style features of two female speakers (0016, 0018) and two male speakers (0013, 0020). As shown in Fig.\ref{tsne}, we observe that the emotional style features form emotional groups for each speaker. These results suggest that these emotional style features characterize well the emotional states. 

We further look into how emotional style features differ between speakers. We measure the similarity of the emotional style features between speaker pairs in terms of euclidean distance and root mean square error (RMSE). 
From Fig.\ref{fig:unique}, we observe the euclidean distance and RMSE within a speaker are lower than those between two speakers, which indicates that emotional style features from SER carry speak-dependent information.

The above analysis shows that we may use emotional style features to encode both speaker-dependent and speaker-independent emotional styles in expressive voice conversion.


\section{A joint speaker identity and emotion style transfer framework}
\vspace{-2mm}
\label{framework}
 We propose a joint emotional style and speaker identity conversion framework using StarGAN, which is referred to as \textit{JES-StarGAN}. We next discuss JES-StarGAN in three stages: 1) emotional style descriptor training, 2) StarGAN training, and 3) run-time conversion. In stage I, we train an auxiliary SER network to act as an emotion descriptor. In stage II, we train a StarGAN network to learn the mapping of both spectral features and emotional style features across different speakers. In stage III, JES-StarGAN performs voice conversion towards target speaker identity and emotional style. 
\vspace{-1mm}
\subsection{Stage I: Emotional style descriptor training }
\vspace{-2mm}
\label{ssec:subhead}
As emotional style presents both speaker-dependent and speaker-independent characteristics at the same time, it is insufficient for us to use a discrete representation to represent different emotional styles, such as one-hot emotion label \cite{zhou2021seen}. 
Therefore, we propose to use deep emotional features that learnt from a large emotional speech corpus to describe different emotional styles.

We propose to train an SER model to learn the emotional style features for different speakers with various emotions. The model architecture is the same as that in \cite{chen20183}, as illustrated in Fig.\ref{SER}. The SER network consists of the following layers: 1) a three-dimensional (3-D) CNN layer; 2) a BLSTM layer; 3) an attention layer; and 4) a fully-connected (FC) layer. The Mel-spectrum input is first projected into a fixed size latent representation by the 3-D CNN, which preserves the useful emotional information while reducing the influence of emotional irrelevant elements. The next BLSTM and attention layer then summarizes the temporal input from the prior layer and produces an utterance-level feature $E$ for emotion category prediction.

\vspace{-1mm}
\subsection{Stage II: JES-StarGAN training}
\vspace{-2mm}
\label{sssec:subsubhead}

StarGAN has been widely used for many-to-many voice conversion and does not need parallel data. We propose a StarGAN-based architecture as shown in Fig.\ref{StarGAN}, which consists of three modules: a generator G, a discriminator D, and a domain classifier C. During stage II, the proposed framework learns a feature mapping of speaker identity and speaker-dependent emotional style across different speakers. 

Given the source acoustic feature sequence $x$, target speaker label $c_y$, and target emotional style feature $E_y$, the generator learns to translate the source acoustic feature sequence $x$ to the target domain by conditioning on the emotional style feature $E_y$ and target speaker label $c_y$. The converted acoustic feature $\hat{y}$ can be represented as:
\begin{equation}
    \hat{y} = G(x,{E_y},{c_y})
\end{equation}
where $c_y$ is a one-hot vector to represent each speaker. In this way, the generator jointly learns the speaker identity information and speaker-dependent emotional style information from the input features.


The discriminator D is designed to judge whether the input is real or not, while the classifier C judges whether the input acoustic features belong to the target speaker. The training process of our proposed method is illustrated in Fig.\ref{StarGAN}. The training losses for our proposed method are described as follows: 


1) \textbf{Adversarial loss:} An adversarial loss is applied to train G and D as follows:
\begin{align}
  L_{adv}^G = -E_{x,E_y,c_y}[log D(G(x,E_y,c_y),c_y)]
 \vspace{-4mm}  
\end{align}

\begin{align}
\begin{split}
    L_{adv}^D ={}& -E_{x,c_x}([log(D(x,c_x))]\\&-E_{x,{E_y},c_y}[log (1-D(G(x,E_y,c_y),c_y))]
\end{split} 
\vspace{-2mm}
\end{align}
where $E[\cdot]$ represents the expectation operation, 
$c_x$ and $E_x$ represent the speaker label and the emotional style features of the source speaker respectively. 
During training, D tries to minimize $L_{adv}^D$ and G tries to minimize $L_{adv}^G$. To G, a smaller value of the adversarial loss indicates a higher similarity between the converted speech and the target emotional speech in terms of both speaker similarity and emotional style.


2) \textbf{Domain classification loss:} A domain classification loss is applied to train the C and G, which is defined as: 
\begin{align}
\label{dom_1}
  L_{dom}^C = -E_{x,c_x}[log {p_C(c_x|x)}]
 \vspace{-4mm}
\end{align}
\begin{align}
\label{dom_2}
  L_{dom}^G = -E_{x,E_y,c_y}[log {p_C(c_y|G(x,E_y,c_y)}]
\end{align}
where $p_C$ represents the output probability distribution from C. During training, C learns to classify a real acoustic feature sequence $x$ to its corresponding speaker label $c_x$ by minimizing Equation \ref{dom_1}. G learns to generate acoustic feature sequence $G(x,E_y,c_y)$ with higher classification accuracy for the target domain $c_y$ by minimizing $L_{dom}^G$ in Equation \ref{dom_2}. The additional input $E_y$ in $G(x,E_y,c_y)$ encourages $G(x,E_y,c_y)$ to carry speaker-dependent emotional style information, so that C can more easily classify $G(x,E_y,c_y)$ with high accuracy in target domain. 


3) \textbf{Cycle-consistency loss:} A cycle-consistency loss is proposed to guarantee the consistency of the contextual information between input and output while converting the speaker identity and emotional style. It is defined as:
\begin{align}
  L_{cyc}^G = E_{x,E_y,c_y,E_x,c_x}[\Vert G(G(x,E_y,c_y),E_x,c_x)-x\Vert _1]
\end{align}
Since the emotional style of source speaker is different from that of target speaker, it is necessary to input $E_x$ as an additional condition for the G to make $G(G(x,E_y,c_y),E_x,c_x)$ closer to the source acoustic features.

4) \textbf{Identity mapping loss:} An identity mapping loss is used to preserve the linguistic information between the same source and target speaker when inputting $x$, $E_x$ and $c_x$ for G. It is defined:
\begin{align}
  L_{id}^G = E_{x,E_x,c_x}[\Vert G(x,E_x,c_x)-x\Vert _1]
\end{align}

Both speaker identity $c$ and $E_x$ are speaker-dependent features for the same speaker. $L_{id}^G$ helps G to generate the acoustic feature sequences from the same speaker be consistent by considering these speaker-dependent features.

The full objective functions of our proposed method are given as follows:
\begin{align}
  L_{D} =L_{adv}^D,~ L_{C} =L_{dom}^C
\vspace{-4mm}
\end{align}
\begin{align}
  L_{G} =L_{adv}^G + \lambda_{dom}L_{dom}^G + \lambda_{cyc}L_{cyc}^G + \lambda_{id}L_{id}^G
\end{align}
where $\lambda_{dom}$,$\lambda_{cyc}$ and $\lambda_{id}$ are trade-off factors to control the relevance of the domain classification loss, the cycle consistency loss and the identity mapping loss to the overall adversarial losses.


At training stage, we condition the generator on emotional style encoding  derived  from  a pre-trained SER model from stage I. JES-StarGAN optimizes the distribution of the generated acoustic features to match that of the target features from emotional data. With the additional input emotional style features, JES-StarGAN learns to project both emotional style and speaker identity into the converted speech. 

\vspace{-1mm}
\subsection{Stage III: Run-time conversion}
\vspace{-2mm}
\label{sssec:subsubhead}
During the run-time conversion, we can convert the acoustic feature sequence $x$ of an input utterance: $\hat{y} = G(x,{E_y'},c_y)$. Since the real target emotion style features are not available from stage I, 
we calculate the mean of emotion style features collected from the corresponding reference utterances as ${E_y'}$, which can be defined as ${E_y'} = mean({E_{y_e}^r})$, where $y_e$ represents target speaker with a certain emotion state.

\section{Experiments}
\label{EXP}

We conduct objective and subjective evaluations to assess the performance of our proposed framework in terms of speaker identity and emotional style. We use a multi-speaker emotional speech dataset, ESD \cite{zhou2021emotional}, to conduct all the experiments. ESD consists of multi-lingual and multi-speaker parallel emotional speech data with five emotions (neutral, happy, sad, angry and surprise), and has been used in emotional voice conversion \cite{zhou2021seen, zhou2021limited} and emotional text-to-speech \cite{liu2021reinforcement}.

We randomly choose three emotions (neutral, happy and sad) and four speakers (two male and two female) from ESD. For each speaker and each emotion, we use 300 utterances for training, 20 utterances for evaluation, and the rest 30 utterances are used as the reference set. 
As a comparative study, we choose StarGAN-VC \cite{kameoka2018stargan} as the baseline.

\begin{table*}[t]
\centering
\scalebox{1}{
\begin{minipage}{\columnwidth}
  \caption{Average MCD [dB] values for 2 male and 2 female speakers with an intra-gender setting.}
  \label{Intra}
  \centering
  \begin{tabular}{c|c||c|c}
  \bottomrule
  \multicolumn{2}{c||}{  }& StarGAN-VC &JES-StarGAN\\
  \hline
  \multirow{2}*{Neutral}&M-M&6.187&5.419\\
  \cline{2-4}
  &F-F&5.917&5.928\\
  \hline
  \multirow{2}*{Happy}&M-M&6.656&6.534\\
  \cline{2-4}
  &F-F&6.190&6.105\\
  \hline
  \multirow{2}*{Sad}&M-M&6.739&6.333\\
  \cline{2-4}
  &F-F&7.240&7.113\\
  \toprule
  \end{tabular}
  \end{minipage}
  
\begin{minipage}{\columnwidth}
  \caption{Average MCD [dB] values for 2 male and 2 female speakers with an inter-gender setting.}
  \label{Inter}
  \centering
  \scalebox{1}{
  \begin{tabular}{c|c||c|c}
  \bottomrule
  \multicolumn{2}{c||}{  }& StarGAN-VC&JES-StarGAN\\
  \hline
  \multirow{2}*{Neutral}&M-F&7.646&7.444\\
  \cline{2-4}
  &F-M&7.655&7.023\\
  \hline
  \multirow{2}*{Happy}&M-F&7.051&6.698\\
  \cline{2-4}
  &F-M&7.479&7.185\\
  \hline
  \multirow{2}*{Sad}&M-F&8.227&7.860\\
  \cline{2-4}
  &F-M&8.222&7.906\\
  \toprule
  \end{tabular}}
  \end{minipage}}
  \vspace{-5mm}
\end{table*}

\subsection{Experimental setup}

\label{sssec:subsubhead}
All the speech data is sampled at 16 kHz and encoded with 16 bits. We extract 36-dimensional Mel-cepstral coefficients (MCEPs), fundamental frequency (F0), and aperiodicity (APs) every 5 ms using WORLD vocoder \cite{morise2016world}. 
36-dimensional MCEPs are used as spectral features, F0 is converted through the logarithm Gaussian (LG) normalized transformation \cite{liu2007high}, and APs are directly copied from the source without any modifications. 

We first train the SER network with IEMOCAP dataset \cite{busso2008iemocap} and then fine-tune it with ESD. We follow the model architecture and training configuration in \cite{chen20183} for SER training. We obtain the 64-dimensional emotional style features as in Fig.\ref{SER}. We then merge the emotion style features together with 36-dimensional MCEPs using a full-connected layer, which is later used as the input to the generator. 

The proposed JES-StarGAN has the following architecture: G consist of an encoder and a decoder.
The encoder consists of 5 layers of CNN, and each is followed by a batch normalization layer and a gated linear unit. The output channel of the encoder is $\{64,128,256,128,10\}$. The decoder consists of 4 layers of CNN followed by a batch normalization layer and a gated linear unit, and a transposed convolution layer. Its output channel is $\{64,128,64,32\}$. 
D consists of four CNN layers (each layer is followed by a batch normalization layer and a gated linear unit), a CNN layer, a sigmoid layer and product pooling layers. The output channel for D is $\{32,32,32,32,1\}$. C consists of a slice layer, four CNN layers (each layer is followed by a batch normalization layer and a gated linear unit), a CNN layer, a softmax layer and product pooling layers. The output channel of C is $\{8,16,32,16\}$.

\begin{table}[t]
\vspace{-3mm}
\centering
    \caption{Mean opinion score (MOS) results, where 15 groups of utterances are evaluated by 13 subjects.}
    \begin{tabular}{c||c}
    \hline 
    Framework & Mean Opinion Score\\
    \hline
    StarGAN-VC& 2.898 $\pm$ 0.36\\
    \hline
    JES-StarGAN&3.044 $\pm$ 0.38\\
    \hline
    \end{tabular}
    \label{MOS}
    \vspace{-5mm}
\end{table}

During training, JES-StarGAN is trained using ADAM optimizer with a learning rate of 0.0001. The batch size is 4 and the training process takes 200k iterations. We set $\lambda_{dom}$=2, $\lambda_{cyc}$=10 and $\lambda_{id}$ = 5.

\subsection{Objective evaluation}

\label{sssec:subsubhead}
We conduct objective evaluation to asses the performance of our proposed framework for neutral, happy and sad utterances. We calculate Mel-cepstral distortion (MCD) \cite{kubichek1993mel} to measure the spectral distortion between the converted and target speech. A smaller value of MCD indicates a smaller spectral distortion and a better conversion performance. 

We first report the MCD results for intra-gender combinations in Table \ref{Intra}: 1) from male to male (denoted as \textsl{M-M}); 2) from female to female (denoted as \textsl{F-F}), and then report the MCD for inter-gender in  Table \ref{Inter}: 3) from male to female (denoted as \textsl{M-F}) and 4) from female to male (denoted as \textsl{F-M}).  
From these results, we observe that the proposed method JES-StarGAN consistently outperforms the baseline StarGAN-VC in both inter-gender and intra-gender conversion, which indicates the effectiveness of our proposed framework. 

\vspace{-3mm}
\subsection{Subjective evaluation}

\label{sssec:subsubhead}
We conduct three listening tests to assess speech quality, speaker similarity and emotional style similarity. 13 subjects participate in all listening tests, in which each listens to 90 converted utterances in total. 

We first report the mean opinion score (MOS) results to evaluate the speech quality. 15 sentences are randomly selected from the evaluation set.  A higher MOS score indicates better speech quality. As shown in Table \ref{MOS}, 
JES-StarGAN outperforms the StarGAN-VC baseline.  

\begin{figure}[t]
    \centering
    \includegraphics[width=5.5cm]{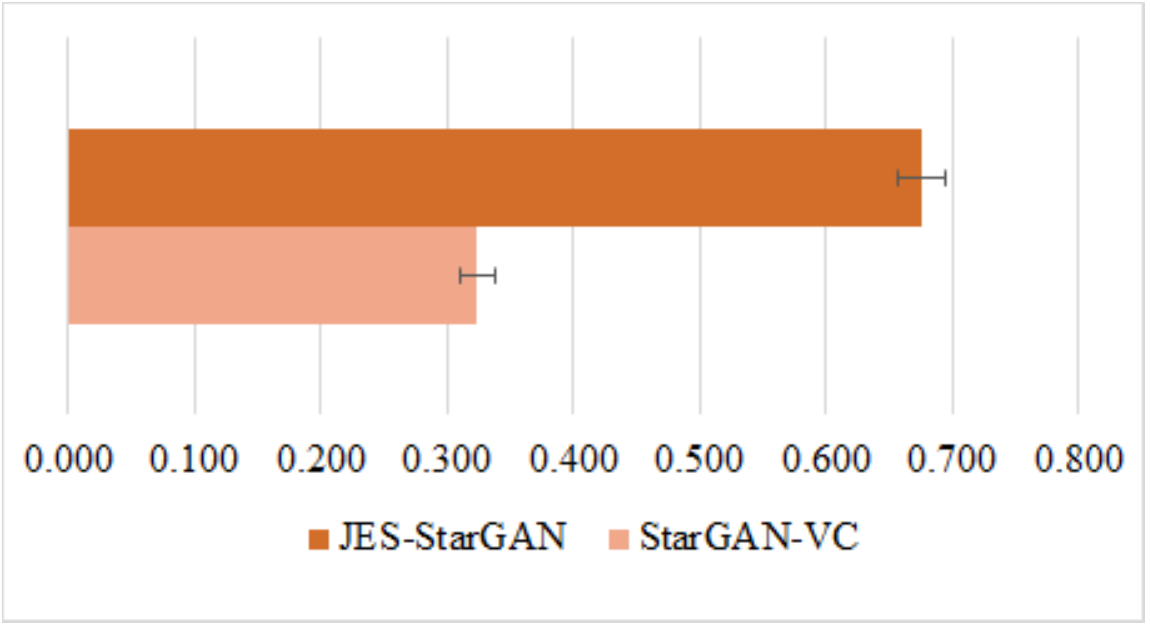}
    \vspace{-3mm}
    \caption{ABX preference results for speaker similarity with 95\% confidence interval, where 15 groups of utterances are evaluated by 13 subjects.}
    \label{fig:ss}
    \vspace{-4mm}
\end{figure}

We further conduct two ABX preference tests to evaluate speaker similarity and emotional style similarity respectively. 
First, we report the results for ABX tests for speaker similarity, where all the subjects are asked to choose the one which sounds closer to the the reference target speech samples in terms of the speaker similarity. As shown in Fig.\ref{fig:ss}, we observe that the proposed JES-StarGAN significantly outperforms the baseline StarGAN-VC in speaker similarity. Benefiting from the joint transfer of speaker identity and emotional style, the proposed JES-StarGAN has a much better performance on the speaker identity conversion than the baseline, which further validates our idea on expressive voice conversion. 

\begin{figure}
    \centering
    \includegraphics[width=5.5cm]{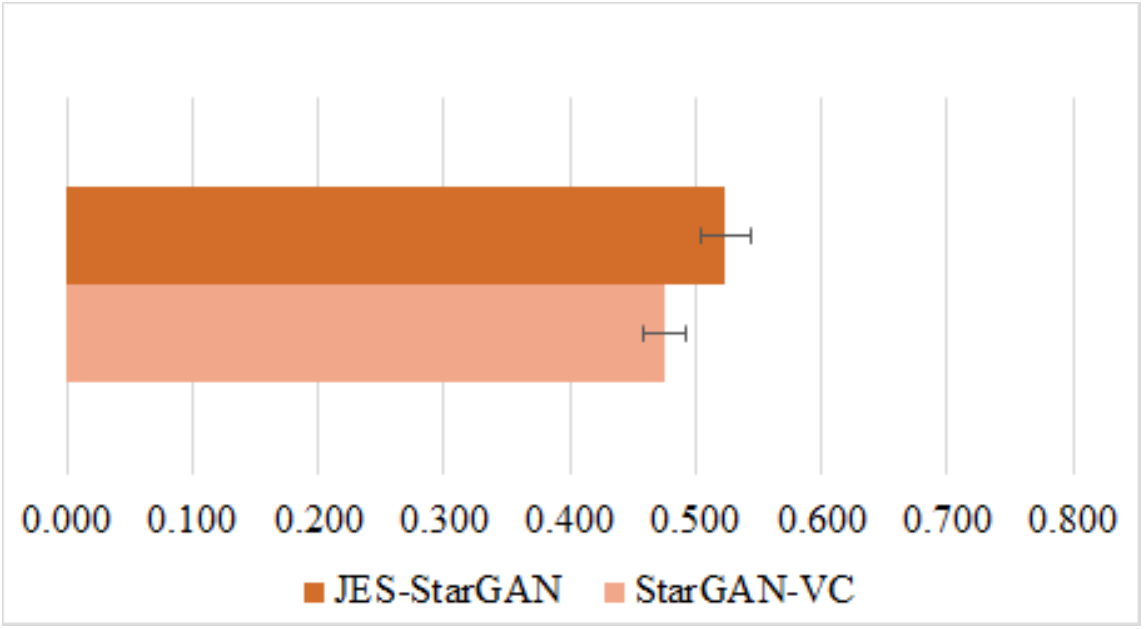}
       \vspace{-4mm}
    \caption{ABX preference results for emotional style similarity with 95\% confidence interval, where 15 groups of utterances are evaluated by 13 subjects.}
    \label{fig:ess}
    \vspace{-4mm}
\end{figure}

We then report the ABX test results for emotional style similarity, where all the subjects are asked to choose the one which sounds closer to the reference target speech samples in terms of the emotional style. As shown in Fig.\ref{fig:ess}, the proposed JES-StarGAN still outperforms the baseline StarGAN-VC in terms of emotional style similarity. It shows the effectiveness of the deep emotional features on emotional style transfer.


\section{Conclusions}
\vspace{-2mm}
\label{sec:page}
This paper marks as the first study for expressive voice conversion. We formulate the problem of expressive voice conversion and  propose a novel solution based on StarGAN to jointly transfer speaker identity and speaker-dependent emotional style without the need for parallel data. We propose to use deep emotional features from SER to characterize different emotional styles in a continuous space. By conditioning the generator with deep emotional features, the framework jointly learn a mapping of speaker-dependent features across different speakers. Experiments show that the proposed JES-StarGAN consistently outperforms the baseline. 

\label{sec:illust}

\section{Acknowledgment}
The research is funded by SUTD Start-up Grant Artificial Intelligence for Human Voice Conversion (SRG ISTD 2020 158) and SUTD AI Grant - Thrust 2 Discovery by AI (SGPAIRS1821).

\bibliographystyle{IEEEbib}
\footnotesize
\bibliography{references}

\end{document}